\documentclass[a4paper,fleqn,usenatbib]{mnras}

\usepackage[T1]{fontenc}
\usepackage{ae,aecompl}

\usepackage{graphicx}   
\usepackage{amsmath}    
\usepackage{amssymb}    

\newcommand{\vsini}{$v \sin i$ }

\usepackage{multirow}
\usepackage{booktabs}

\title[Interstellar C$_{60}^+$]{C$_{60}^+$ - looking for the bucky-ball in interstellar space}

\author[G.A. Galazutdinov et al.]{
G.A. Galazutdinov$^{1,2,3},$\thanks{E-mail: runizag@gmail.com}
V.V. Shimansky$^{4}$,
A. Bondar$^5$,
G. Valyavin$^3$,
J. Kre{\l}owski$^6$
\\
$^{1}$Instituto de Astronomia, Universidad Catolica del Norte, Av. Angamos 0610, Antofagasta, Chile\\
$^{2}$Pulkovo Observatory, Pulkovskoe Shosse 65, Saint-Petersburg 196140, Russia\\
$^{3}$Special Astrophysical Observatory, Nizhnij Arkhyz, 369167, Russia\\
$^{4}$Kazan Federal University, Kazan, 420008, Russia\\
$^{5}$International Center for Astronomical and Medico-Ecological Research, Zabolotnoho Str. 27, Kiev, 03187, Ukraine \\
$^{6}$Center for Astronomy, Nicholas Copernicus University, Gagarina 11, Pl-87-100 Torun, Poland
}

\date{Accepted XXX. Received YYY; in original form ZZZ}

\pubyear{2016}

\begin{document}

\label{firstpage}
\pagerange{\pageref{firstpage}--\pageref{lastpage}}
\maketitle

\begin{abstract}
The laboratory gas phase spectrum  recently published by Campbell et
al. has reinvigorated attempts to confirm the presence of the
C$_{60}^+$ cation in the interstellar medium, thorough an analysis of the
spectra of hot, reddened stars. This search is hindered by at
least two issues that need to be addressed: (i) the
wavelength range of interest is severely polluted by strong water-
vapour lines coming from the Earth's atmosphere; (ii) one of the
major bands attributed to C$_{60}^+$, at 9633 \AA, is blended
with the stellar Mg{\sc ii} line, which is susceptible to non-local-thermodynamic equilibrium  effects
in hot stellar  atmospheres. Both these issues are here carefully
considered here for the first time, based on  high-resolution and
high signal-to-noise ratio echell\'e spectra for  19 lines of
sight. The result is that the presence of C$_{60}^+$  in interstellar
clouds is brought into question.
\end{abstract}



\section{Introduction}

One of the carbon allotropes, the spherical fullerene molecule
C$_{60}^+$ (buckminsterfullerene, bucky-ball) is of particular
interest for astrophysics. Indeed, in their seminal paper
presenting the discovery of C$_{60}^+$ (Kroto et al., 1985),
authors suggested that the molecule might give rise to a super
stable species which might exist in interstellar space and
circumstellar shells. A decade later, the first attempt to
find C$_{60}^{+}$ in interstellar medium was made by Foing \&
Ehrenfreud (1994), who reported on the detection of two
interstellar absorptions, at 9577 and 9633 \AA, almost
coincident with the wavelengths expected for C$_{60}^{+}$.  However, the limited
number of observed sight lines, and the fact  that the laboratory
spectra were acquired in solid state matrices, made that
identification uncertain.  Jenniskens et al. (1997) and
Galazutdinov et al. (2000) pointed out two difficulties inherent
to the search for interstellar C$_{60}^{+}$ (see the Abstract
above), and mentioned the non-detection of minor 
C$_{60}^{+}$ bands, expected at 9429 and 9366 \AA\ (telluric lines
are particularly strong in this wavelength range). Galazutdinov et
al. (2000) also presented the first,  simplified attempt to
eliminate the influence of stellar Mg{\sc ii} line.

In 2010 Cami et al. reported on the first detection of infrared
emissions carried by neutral C$_{60}$ seemingly attached to the
solid material (dust particles). These emissions were observed  in the
vicinity of a peculiar planetary nebula Tc1. Independently,
Sellgren et al. (2010) reported on the presence of neutral
C$_{60}$ in the NGC 7023 reflection nebula illuminated by a B star
HD\,200775. The same nebula was recently identified  as a source
of C$_{60}^+$, following the discovery of  emission bands at 6.4,
7.1, 8.2, and 10.5 \(\mu\)m by Bern\'e et al. (2013). The molecule
was detected only  in  the regions closest to the star. All these
reports fostered the hope that the
buckminsterfullerene cation would be detected in translucent interstellar clouds.
However, the laboratory   spectra of  C$_{60}^+$, available in 1994,
were measured for a solid medium (cryogenic matrix), and these spectra
therefore could not offer any firm confirmation of this ion as the one
responsible for known  9577 and 9633 \AA\ diffuse bands.

The dormant interest in discovering interstellar bucky-ball
cations was revived following the recent publication of laboratory
gas-phase spectra by Campbell et al. (2015). According to the
authors, C$_{60}^{+}$  exhibits four relatively strong spectral
lines, centered at 9365.9$\pm$0.1, 9428.5$\pm$0.1, 9577.5$\pm$0.1
and 9632.7$\pm$0.1 \AA, with relative intensities of 0.2, 0.3,
1.0 and 0.8 respectively. New attempts to confirm the presence of
this molecule in the interstellar medium, by analysing the spectra of hot,
reddened stars, were recently reported by Walker et al.
(2015), who have also tried to detect the two weaker spectral
features of C$_{60}^{+}$ in the sight lines towards HD\,183143 and
HD\,169454. Very recently, Campbell, Holz \& Maier (2016) reported on the
detection of a very weak feature at 9348.5 \AA\  (just 1/10 of the
intensity of the major 9577 \AA\ band), again in a spectrum of
HD\,183143. Unfortunately, both these studies lack an analysis
of possible pollution by telluric lines. Indeed, as we
demonstrate below, the DIB9633 in both HD\,169454 and
HD\,183143 spectra significantly contaminated by stellar Mg{\sc
ii} 9632\AA\ line. Generally, the HD\,183143 sight line is not a good
choice for finding any new diffuse bands. As demonstrated in the
respective DIB surveys  (Galazutdinov et al. 2000b, Hobbs et al.
2009), it is often not easy to distinguish between stellar and
interstellar lines towards this relatively cool and slowly
rotating white hypergiant classified as B6 to B8 Ia-0 (Chentsov
2004).

Here we present the most careful analysis to date of near-infrared
interstellar features commonly attributed to C$_{60}^{+}$, for a
relatively large sample of targets. We took into account the
possibilities of spectral contamination not only from telluric
but also from stellar lines. Indeed, incidental overlap with
stellar lines is at particularly important danger in cases such as
HD\,183143, where  broad interstellar features have widths
comparable to the stellar lines, the latter being numerous in this
late B type supergiant.

\section{Spectral data}

Our sample of astronomical data includes precise measurements of
the 9577 \AA\ and 9633 \AA\ bands,  the  two major interstellar
features allegedly resulting from C$_{60}^{+}$. High signal-to-noise and
high resolution echell\'e spectra were obtained for 19 heavily
reddened targets (Tables 1 and 2) with the UVES spectrograph fed
by the Kueyen 8-m telescope at Paranal. The resolving power
R=$\lambda$/$\Delta\lambda$ was 80,000 in the range of Mg{\sc ii}
4481.2 \AA\ line, and 110,000 in the range of currently analysed
diffuse interstellar bands (DIB).

All spectra were processed and measured in a standard way using
both IRAF (Tody 1986) and our own
DECH\footnote[1]{http://gazinur.com/DECH-software.html} codes. The
wavelength scale of all spectra was corrected by well-known
atomic/molecular interstellar lines, e.g. K{\sc i} and C$_2$.

\begin{table*}
    \caption{Basic parameters of the observed stars. Effective temperature - T$_{eff}$ (K),
    logarithm of gravity - $\lg\,g$ ($cm\,s^{-2}$), microturbulent velocity - v$_{turb}$ ($km\,s^{-1}$),
    projected rotational velocity - v\,sin$i$ ($km\,s^{-1}$), abundances of marked chemical elements
    relative to the solar abundance - [X/H] (dex).  
    }
    \label{stars}
\centering{
\begin{tabular}{rrlccccccccccc}
\hline
\multirow{2}{*}{Star}
           &\multirow{2}{*}{T$_{eff}$}
                   &\multirow{2}{*}{$\lg\,g$}
                          &\multirow{2}{*}{v$_{turb}$}
                               &\multirow{2}{*}{v\,sin$i$}
                                     &  \multicolumn{9}{c}{Abundance [X/H]}                                   \\
\cmidrule{6-14}
          &        &      &    &     &  He   &   C   &    N   &  O    &  Fe   & Si    &  Mg   &   Al  &  Ne   \\
\hline
CD-324348 & 19,500 & 2.45 & 9  & 36  & +0.30 & -0.30 &  +0.30 & -0.30 & -0.20 & +0.50 & +0.30 & -0.30 & +0.20 \\
BD-145037 & 18,000 & 1.80 & 10 & 42  & +0.30 & +0.10 &  +0.30 & +0.10 & +0.20 & +0.30 & +0.10 & +0.20 &       \\
HD 23180  & 24,000 & 3.45 & 13 & 78  & +0.15 & -0.10 &  -0.20 & -0.10 & -0.20 & -0.40 & -0.25 &       &       \\
HD 27778  & 15,500 & 3.80 & 5  & 92  & +0.20 & -0.05 &  -0.20 & +0.05 & -0.40 & -0.40 & -0.50 &       &       \\
HD 63804* & 9,400? & 1.10?& 6? & ?   & +0.15:&-0.20: &        &       & -0.20:&       & -0.30:&       &       \\
HD 76341  & 34,000 & 3.70 & 13 & 66  & +0.25 & +0.20 &  +0.40 & +0.20 &       & +0.40 & -0.05 & -0.10 & +0.20 \\
HD 78344  & 31,000 & 3.30 & 13 & 98  & +0.30 & +0.30 &  +0.50 & +0.30 &       & +0.50 & +0.10 & +0.10 & +0.30 \\
HD 80077  & 17,000 & 2.00 & 15 & 47  & +0.05 & -0.40 &  +0.40 & +0.10 & +0.10 & +0.40 & -0.05 & +0.10 &       \\
HD136239  & 17,000 & 1.80 & 11 & 43  & +0.40 & +0.30 &  +0.50 & +0.30 & +0.20 & +0.50 & +0.30 & +0.20 &       \\
HD145502  & 21,000 & 4.00 & 8  & 98  & -0.10 & -0.70 &  +0.40 & +0.05 & -0.20 & +0.15 & -0.60 &       &       \\
HD147888  & 16,000 & 4.10 & 5  & 104 & -0.10 & -0.10 &  -0.10 & +0.00 & -0.30 & -0.10 & -0.50 &       &       \\
HD148379  & 17,000 & 1.70 & 14 & 51  & +0.30 & -0.10 &  +0.30 & +0.10 & +0.10 & +0.20 & +0.10 & +0.00 &       \\
HD148605  & 20,500 & 4.20 & 11 & 145 & +0.15 & -0.30 &  -0.40 & -0.30 & -0.60 & -0.60 & -0.60 &       &       \\
HD167264  & 29,000 & 3.20 & 16 & 82  & +0.35 & +0.10 &  +0.30 & +0.25 & -0.20 & -0.10 & +0.00 &       &       \\
HD168625  & 14,000 & 2.00 & 10 & 52  & +0.00 & -0.20 &  +0.25 & -0.05 & -0.10 & +0.05 & +0.05 & +0.00 &       \\
HD169454  & 21,000 & 2.10 & 16 & 39  & +0.10 & +0.30 &  +0.50 & +0.30 &       & +0.60 & +0.30 & +0.20 & +0.20 \\
HD170740  & 21,000 & 3.90 & 10 & 40  & +0.00 & -0.05 &  -0.15 & -0.20 & -0.40 & -0.40 & -0.45 &       &       \\
HD183143  & 11,500 & 1.40 & 8  & 37  & +0.00 & +0.00 &  +0.30 & +0.10 & -0.20 & +0.10 & +0.10 &       &       \\
HD184915  & 27,000 & 3.40 & 19 & 220 & +0.30 & +0.10 &  +0.50 & +0.00 &       & -0.10 & +0.00 &       &       \\
\hline
    \end{tabular}
}
\\
$^*$ - see the comment in  "Mg{\sc ii} contamination" section.
\end{table*}

\begin{table*}
        \caption{Measurements of four C$_{60}^+$ candidate bands. Upper, cumulative column captions give: laboratory central wavelength, laboratory full width at half maximum (FWHM), and
laboratory intensity normalized to the strength of the 9577~\AA\
band -- (I/I$_{9577}$)$_{lab}$ (Campbell et al., 2015). The
lower captions give: observed central wavelength (\AA),
observed equivalent widths (m\AA), observed FWHM, and intensity
normalized to that of the 9577~\AA\ feature. Two  equivalent
width values are given for the 9633~\AA\ band: EW$_o$ (before
correction) and EW$_c$ (after the removal of the Mg{\sc ii} blending
feature). Measurement errors are given in parentheses.
\newline
{\it Continued...}}
    \label{c60data}
\centering{
\begin{tabular}{lccccccccc}
\hline

\multirow{2}{*}{Star}
          & \multicolumn{5}{c}{9632.7(0.1), 2.2(0.2), (I/I$_{9577}$)$_{lab}$=0.8}
                                                            & &\multicolumn{3}{c}{9577.5(0.1), 2.5(0.2), I$_{lab}$=1.0}
                                                                                               \\
\cmidrule{2-6}  \cmidrule{8-10}
            &$\lambda_c$  & EW$_o$(m\AA)
                                   & EW$_c$(m\AA)
                                            &  FWHM(\AA) &I/I$_{9577}$
                                                             & &$\lambda_c$& EW$_o$(m\AA)
                                                                                     & FWHM(\AA)  \\  
\hline
CD-32 4348  &9632.2 (0.1) & 173(17)&  75(16)& 2.1(0.1) & 0.4 & &9576.9(0.2)& 168(11) & 3.0(0.3) \\  
BD-14 5037  &9632.5 (0.1) & 202(20)& 125(15)& 2.9(0.3) & 1.2 & &9577.1(0.2)& 107(10) & 2.8(0.2) \\  
HD 23180    &9633.1 (0.2) & 145(40)& 141(46)& 3.7(0.3) & 1.8 & &9577.5(0.2)&  77(34) & 3.6(0.3) \\  
HD 27778    &9632.6 (0.2) &  94(22)&  64(17)& 2.7(0.2) & 1.3 & &9577.5(0.2)&  50(30) & 2.2(0.3) \\  
HD 63804$^*$&9632.2 (0.3) & 150(24)&  20(20)& 2.7(0.3) & 0.1 & &9576.9(0.1)& 207(11) &          \\  
HD 76341    &9632.2 (0.2) & 136(30)& 134(30)& 2.5(0.2) & 1.2 & &9577.2(0.2)& 110(20) & 2.5(0.2) \\  
HD 78344    &9632.3 (0.2) & 170(13)& 170(13)& 2.9(0.2) & 0.6 & &9576.9(0.2)& 294(15) & 3.7(0.3) \\  
HD 80077    &9632.15(0.15)& 168(14)&  95(11)& 2.0(0.2) & 0.6 & &9577.3(0.1)& 160(10) & 2.9(0.2) \\  
HD136239    &9631.8 (0.1) & 252(25)& 120(20)& 2.4(0.2) & 0.6 & &9576.9(0.2)& 195(15) & 2.9(0.2) \\  
HD145502    &9632.3 (0.2) & 160(20)& 158(20)& 3.1(0.2) & 1.3 & &9576.9(0.2)& 120(30) & 3.5(0.3) \\  
HD147888    &9632.2 (0.2) & 175(11)& 110(12)& 3.2(0.2) & 1.6 & &9576.9(0.2)&  70(20) & 3.3(0.3) \\  
HD148379    &9632.2 (0.2) & 182(11)&  80(11)& 2.4(0.2) & 0.6 & &9577.3(0.2)& 137(9)  & 3.3(0.2) \\  
HD148605    &9632.25(0.1) & 120(30)& 119(35)& 3.4(0.2) & 1.5 & &9576.9)0.2)&  80(20) & 3.7(0.4) \\  
HD167264    &9632.4 (0.2) &  88(18)&  82(20)& 2.7(0.2) & 1.4 & &9576.8(0.2)&  60(17) & 3.2(0.2) \\  
HD168625    &9631.5 (0.1) & 342(26)& 194(25)& 1.8(0.1) & 0.6 & &9576.2(0.1)& 320(25) & 3.1(0.2) \\  
HD169454    &9631.4 (0.3) & 210(25)& 130(20)& 2.7(0.2) & 1.6 & &9577.1(0.1)&  82(10) & 2.3(0.2) \\  
HD170740    &9632.1 (0.2) & 175(20)& 150(20)& 2.8(0.1) & 1.6 & &9576.9(0.2)&  93(20) & 3.0(0.4) \\  
HD183143    &9632.5 (0.2) & 230(20)& 105(20)& 1.9(0.3) & 0.4 & &9577.3(0.2)& 300(20) & 2.9(0.2) \\  
HD184915    &9632.4 (0.3) &  71(25)&  70(25)& 2.5(0.1) & 1.0 & &9576.9(0.2)&  70(20) & 3.5(0.3) \\  
\hline
\end{tabular}
}
\\
$^*$ - The corrected equivalent width of DIB9633 measured towards HD\,63804 is  an upper limit, owing to uncertain basic stellar parameters.
\end{table*}

\begin{table*}
        \contcaption{.}
    \label{c60data:continued}
\centering{
\begin{tabular}{lccccccccc}
\hline
\multirow{2}{*}{Star}
          & \multicolumn{4}{c}{9428.5(0.1), 2.4(0.1), (I/I$_{9577}$)$_{lab}$=0.3}
                                                       && \multicolumn{4}{c}{9365.9(0.1), 2.4(0.1), (I/I$_{9577}$)$_{lab}$=0.2} \\
\cmidrule{2-5}  \cmidrule{7-10}
          & $\lambda_c$ & EW$_o$(m\AA)
                                  & FWHM(\AA)  &I/I$_{9577}$ && $\lambda_c$ & EW$_o$(m\AA) & FWHM(\AA)  &I/I$_{9577}$\\
\hline
CD-32 4348 &     n/a     & $<$5    &          & $<$0.1   &&               \multicolumn{4}{c}{n/a}          \\
BD-14 5037 &   \multicolumn{4}{c}{strong telluric}       && 9365.4(0.3) & 30(30)       & 2.2(0.2) & 0.15   \\
HD 23180  &      \multicolumn{4}{c}{n/a}                &&          \multicolumn{4}{c}{n/a}                \\
HD 27778  &     \multicolumn{4}{c}{n/a}                 &&             & $<$50        &          & $<$1   \\
HD 63804  &    \multicolumn{4}{c}{n/a}                  &&                \multicolumn{4}{c}{n/a}          \\
HD 76341  &      \multicolumn{4}{c}{n/a}                &&             & $<$30(30)    &          & $<$0.3 \\
HD 78344  &              & $<$2(2)&          & $<$0.01  &&             & $<$30830)    &          & $<$0.1  \\
HD 80077  & 9428.15(0.1) & 50(50) & 2.5(0.5) &   0.3    && 9365.5(0.3) & 50(50)       & 2.5(0.5) & 0.3    \\
HD136239  &      \multicolumn{4}{c}{strong telluric}   &&    \multicolumn{4}{c}{strong telluric}          \\
HD145502  &      \multicolumn{4}{c}{n/a}                &&    \multicolumn{4}{c}{n/a}                      \\
HD147888  &      \multicolumn{4}{c}{n/a}                &&    \multicolumn{4}{c}{n/a}                      \\
HD148379  &      \multicolumn{4}{c}{n/a}                && 9365.7(0.3) & 38(25)       & 2.6(0.3) & 0.2     \\
HD148605  &      \multicolumn{4}{c}{n/a}                &&    \multicolumn{4}{c}{n/a}                      \\
HD167264  &      \multicolumn{4}{c}{n/a}                &&    \multicolumn{4}{c}{n/a}                      \\
HD168625  &     \multicolumn{4}{c}{n/a}                 &&             & $<$60        &          & $<$0.25 \\
HD169454  &     \multicolumn{4}{c}{n/a}                 &&    \multicolumn{4}{c}{strong telluric}          \\
HD170740  &      \multicolumn{4}{c}{n/a}                &&    \multicolumn{4}{c}{n/a}                      \\
HD183143  &              & $<$50   &         & $<$0.1   &&    \multicolumn{4}{c}{strong telluric}          \\
HD184915  &      \multicolumn{4}{c}{n/a}                &&   \multicolumn{4}{c}{n/a}                       \\
\hline
\end{tabular}
}
\end{table*}

\begin{table}
    \caption{Oscillator strength of 4481 and 9632 \AA\ Mg{\sc ii} triplets.}
    \label{Mg2gf}
\centering{
\begin{tabular}{rrrr}
\hline
$\lambda$,(\AA)&  lg gf &$\lambda$,(\AA)&    lg gf  \\
\hline
  4481.1260    &  0.74   &  9631.8910    &  0.59      \\
  4481.1500    & -0.56   &  9631.9470    & -0.71      \\
  4481.3250    &  0.59   &  9632.4300    &  0.43      \\
\hline
\end{tabular}
}
\end{table}

\subsection{Telluric contamination}

\begin{figure}
    \includegraphics[width=4.5 cm, angle=270]{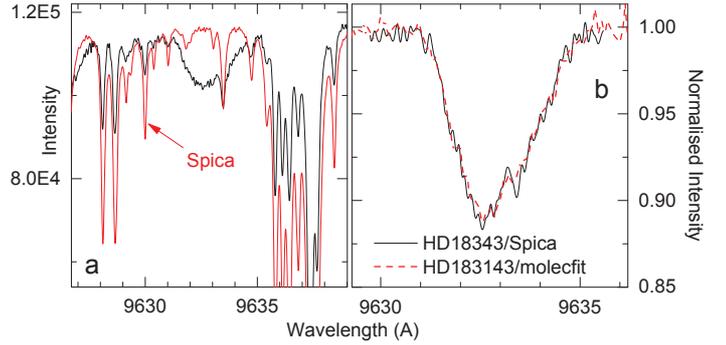}
    \caption{ {\bf a)} Spectra of HD\,183143 and Spica in the vicinity of DIB9633.
     {\bf b)} Comparison of DIB9633 profiles after the removal of telluric lines using Spica and MOLECFIT.}
    \label{molcfit}
\end{figure}

Near infrared wavelength range, when observed using ground--based
instruments, is subject to strong contamination by telluric lines.
In order to eliminate them, we applied the classical method implemented based
on the use of a divisor, namely the spectrum of an unreddened,
hot, and preferably rapidly rotating, star. Our software allowed for the
compensation of both positional and intensity variability within
the set of observed telluric lines. As a divisor we used the
UVES spectra of Spica (HD\,116658). Recently, Tkachenko et al.
(2016) published an extensive study of this close binary system,
which has an orbital period of $\sim$4$^d$, effective temperatures of
25,300$\pm$500 K and 20,900$\pm$500 K, logarithms of gravity
3.71$\pm$0.10 and 4.15$\pm$0.15, rotational velocities \vsini
165.3$\pm$4.5 km/s and 58.8$\pm$1.5 km/s, for the primary and
secondary components respectively. When used for the removal of
telluric lines, the complex and variable spectrum of Spica may
introduce unwanted distortion into relatively broad profiles, such as that of
DIB9633, owing to the  presence of stellar line(s). This is
depicted in Figure \ref{molcfit}a for the DIB9633 profile
observed towards HD\,183143. The lack of a stellar, contaminating
component inside the DIB9633 profile, is evident. However, in order to
check for the applicability of Spica's spectra as divisors, we
tested the recently introduced MOLECFIT procedure (Smette et al.,
2015), a software tool designed to remove atmospheric
absorption features. Fig. \ref{molcfit}b shows a good matching of
the DIB9633 profiles derived  with the divisor (Spica) and
MOLECFIT approaches. The same quality of agreement has also been
found for the other three bands that are candidates for C$_{60}^+$
absorptions. We checked the removal of telluric lines for all
targets using both methods.

\begin{figure}
    \includegraphics[width=6 cm, angle=270]{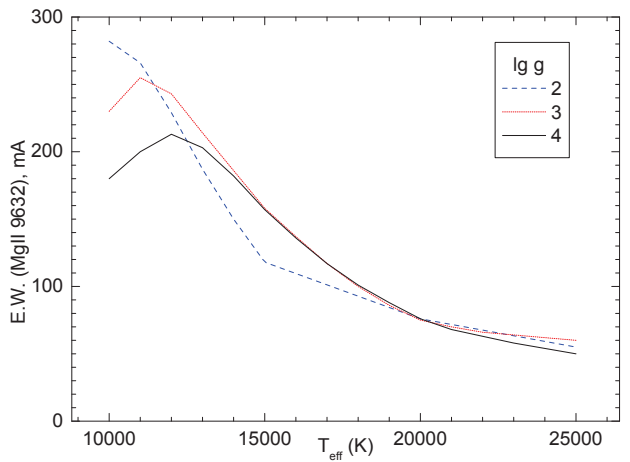} 
    \caption{Variation of the non-LTE equivalent width of Mg{\sc ii} 9632 \AA\ line with the change of stellar parameters.
    Note that model spectra are calculated assuming the solar chemical composition and a fixed  microturbulent
    velocity value, 5 km/s. In general, microturbulent velocity grows with temperature rise and
    with decrease of gravity. }
    \label{Mg2ew}
\end{figure}

\section{Mg{\sc ii} contamination}

\begin{figure}
    \includegraphics[width=6 cm, angle=270]{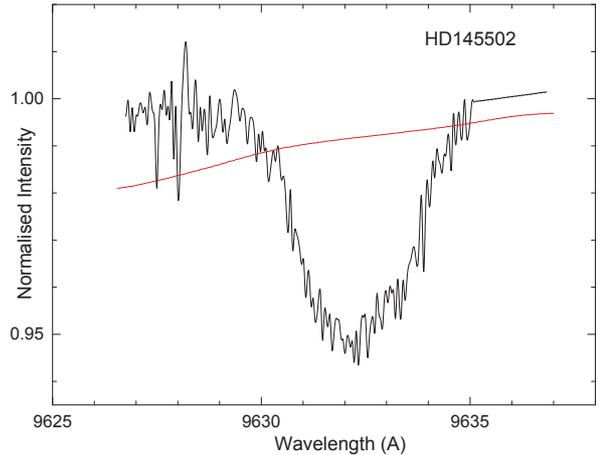} 
    \caption{DIB9633 profile, as observed towards HD 145502, after the telluric correction (the corresponding fragment of  synthetic
     stellar spectrum is shown in red).
This exemplifies the case of no detectable Mg{\sc ii} effect.   }
    \label{shallowMg2}
\end{figure}

Background stars are obviously not ideal sources of radiation for
measuring interstellar absorptions, particularly because of the
presence of their own spectral features. The pollution of the DIB9633
profile by the stellar Mg{\sc ii} line, even though  it has been discussed in
the past, has never been  properly taken into account. Estimations
of Mg{\sc ii} 9632 \AA\ line strengths are complicated by the
sensitivity to non-local thermodynamic equilibrium (non-LTE) effects present in  atmospheres of hot
stars. The variability of this atomic feature with changing
stellar parameters (for solar chemical composition only) is shown
in Fig. \ref{Mg2ew}. However, we caution against a simplified
conclusion that the removal of the Mg{\sc ii} 9632~\AA\ line is
always compulsory. For the majority of hot objects the line is in
fact weak, shallow, and much broader than DIB9633, owing to the rapid
rotation and/or high  gravity (e.g. HD\,145502, see Table 1), so
that it  may not noticeably alter the coinciding DIB profile (see
Fig \ref{shallowMg2}).

\begin{figure}
    \includegraphics[width=10 cm, angle=0]{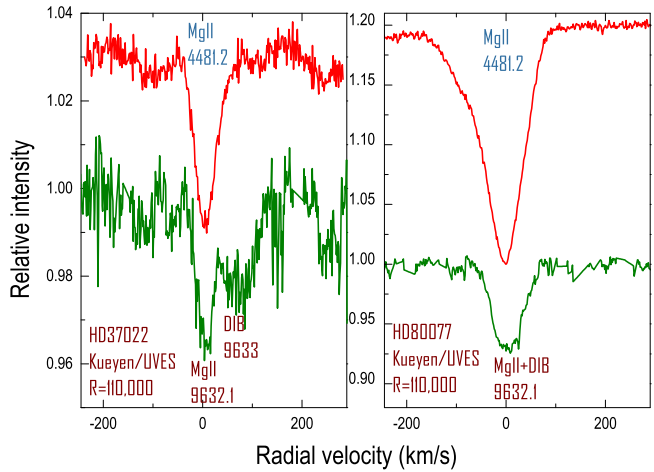} 
    \caption{Stellar Mg{\sc ii} 4481.2 \AA\ line versus the DIB9633 profile, as plotted against the radial velocity scale.
     The presence of  Mg{\sc ii} 9632 \AA\ line alongside the DIB is obvious for HD\,37022. Conversely, both spectral features overlap perfectly in the case of HD\,80077, so that the DIB is much weaker than it appears, and   the separation of two contributions   requires a careful procedure.      }
    \label{Mg2inTwoStars}
\end{figure}

There are particular "lucky" instances where the blend of DIB9633
and Mg{\sc ii} can be easily resolved. This may occur when the
difference between the radial velocity of a background star and
the velocity of the studied interstellar cloud is large or when
the DIB is intrinsically shifted in relation to its ``normal''
positions. HD\,37022 offers the example of this latter possibility
(Fig. \ref{Mg2inTwoStars}) with the red--shift of DIB9633
discovered by Kre{\l}owski \& Greenberg (1999) and recently
confirmed by Kre{\l}owski et al. (2015). On the other hand, even
an evidently strong stellar line may hide completely inside the
DIB profile when the radial velocities of a star and  the cloud
are  similar (Fig. \ref{Mg2inTwoStars}).

\begin{figure}
    \includegraphics[width=4.5 cm, angle=270]{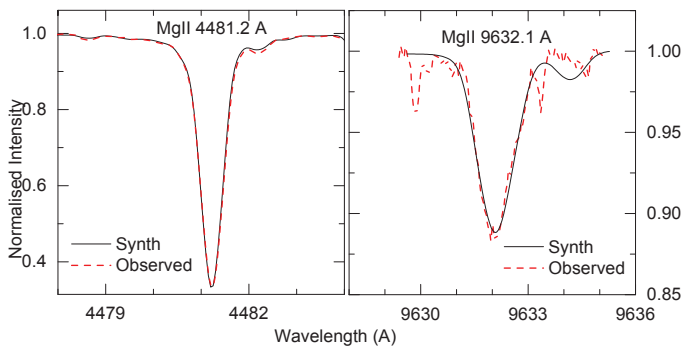} 
    \caption{Profiles of Mg{\sc ii} 4481.2 and 9632.1 \AA\ lines in the spectrum of Sirius, shown with the corresponding synthetic spectrum.
     The corrected oscillator strengths are given in Tab. 3.
     }
    \label{Sirius}
\end{figure}

\begin{figure*}
    \includegraphics[width=6 cm, angle=270]{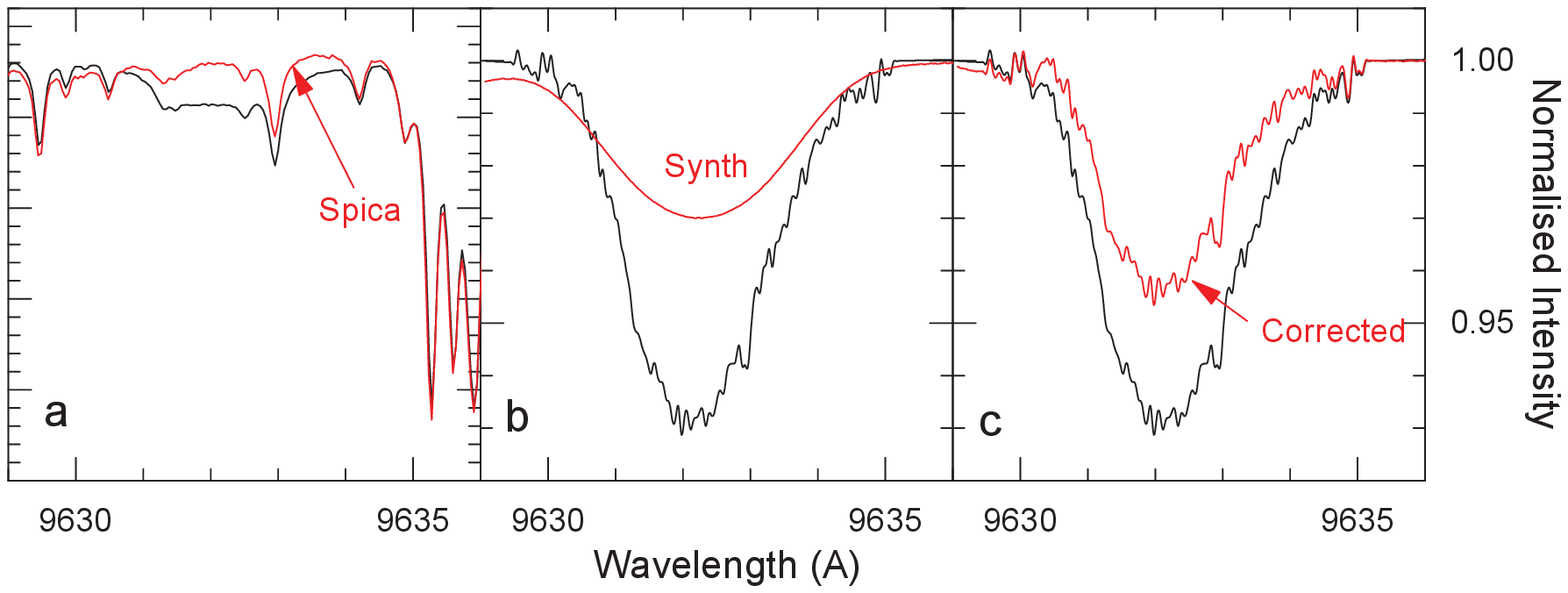} 
    \caption{Consecutive stages of the  DIB9633 profile correction, as illustrated for the HD\,80077 sight line.
    {\bf a)} Initial stage. Original  spectrum,  polluted by strong telluric lines, shown together with the spectrum of Spica, serving as the divisor;
    {\bf b)}  DIB profile with telluric lines removed (black). Smooth red curve represents the synthetic non-LTE
    profile of the stellar Mg{\sc ii} 9632 line, derived with the parameters given in Tab. 1;
    {\bf c)} Final stage. DIB9633 corrected for the presence of  Mg{\sc ii} line, shown together with the uncorrected profile.
    }
    \label{hd80077} 
\end{figure*}


The stellar line of ionized magnesium can thus in many cases
severely contaminate the DIB9633 profile. In order to eliminate this
effect, we first corrected the oscillator strength (gf) of
the Mg{\sc ii} 9632.1~\AA\ transition, using a high quality UVES
spectrum of Sirius, where both neutral and ionized magnesium lines
are easily seen. The magnesium abundance was estimated based on
the well-known lines at 4703.0, 5183.6, 5528.4~\AA\AA\ (Mg{\sc i})
and 4481.2 \AA\ (Mg{\sc ii}) measured in the spectrum of Sirius.
Then, the oscillator strength of the 9632~\AA\ Mg{\sc ii} line
(Table 3) was corrected by matching the observed profile with the
model spectrum (Fig.\ref{Sirius}).

The basic stellar parameters of the program stars (effective
temperature, logarithm of gravity, rotational velocity, chemical
composition) were determined with the aid of STAR code
(Menzhevitski et al., 2014) which takes account of non-LTE effects
for H{\sc i}, He{\sc i}, He{\sc ii}, C{\sc ii}, C{\sc iii} and
Mg{\sc ii} species.  Hydrostatic models of stellar atmospheres
were calculated using the program ATLAS12 (Castelli \& Kurucz,
2004), and the non-LTE populations for the above mentioned
elements were derived using the software package NONLTE3
(Sakhibullin, 1983).

The magnitude of non-LTE effects and chemical composition of the
investigated stars were derived by the analysis of certain atomic
lines located in wavelength ranges free of telluric contamination
and having  well-determined parameters, for example:
H$\varepsilon$-H$_8$, He{\sc i} 4009.2, 4026.2, 4387.9, 4471.5,
4713.2, 4921.9, 5047.8 \AA, He{\sc ii} 4199.8, 4541.6 \AA, C{\sc
ii} 3918.9, 3920.8, 4266.9 \AA, C{\sc iii} 4651.5, 4665.9 \AA,
Mg{\sc ii} 4481.2\AA.

The basic parameters of the observed stars and the derived abundance
of several chemical elements are listed in Table 1.  It is
confirmed that subtraction of the stellar 9632~\AA\ Mg{\sc ii}
line  may essentially reduce the intensity of the interstellar
feature at 9633~\AA\  (Fig. \ref{hd80077}). Conversely, DIB9577 is
not polluted by stellar lines, for all  the sight lines investigated
here. Our results for the four DIBs previously assigned to
C$_{60}^{+}$ are collected in Table 2. Note that equivalent widths
of DIB9633 are given in two separate columns which list the values
without  and with the correction for the stellar Mg{\sc ii} line.

Below we provide  information specific to  each of the studied targets.
\begin{description}
  \item[\bf CD-32 4348]The abundance of alpha elements is [x/H] = -0.4 dex. Moderate deficits of helium, magnesium and nitrogen are revealed,
                  while silicon exhibits
                  an overabundance. A strong blend of Mg{\sc ii} 4481 \AA\ and Al{\sc iii} 4479 \AA\ lines generated in the model
                  spectrum matches the observed profile well. Subtracting the  9632 \AA\ line of Mg{\sc ii}  reduces
                  the intensity of DIB9633 almost half.
                  Minor bands at 9429 and 9366 \AA, expected for C$_{60}^{+}$ are below the level of detection.
  \item[\bf BD-14 5037]Overabundances of He, N, Si (up to +0.3 dex) are typical for stars experiencing the CNO-cycle.
                  The magnesium abundance [Mg/H] = 0.1 dex is accurately estimated by means of the Mg{\sc ii} 4481 \AA\
                  line. This object provides an example  of the diffuse band at 9633 \AA\ being  stronger than that at 9577 \AA,
                  despite that the intensity of former being reduced by the removal of the contribution from stellar magnesium.
  \item[\bf HD 23180] There is a good coincidence of the observed phase of this binary target with the model spectrum.
                  The magnitude of non-LTE effects is low. The chemical composition exhibits a small
                  deficit of heavy elements. The binarity of the target is well seen in the observed spectrum but can be reproduced.
                  The doubled profile of  Mg{\sc ii} 4481 \AA\ line is satisfactorily coincident with the model spectrum.
                  A weak  Mg{\sc ii} line at 9632 \AA\ fades with strong He{\sc i} line to a smeared blend.
  \item[\bf HD 27778] There is good coincidence of observed and model spectra. The magnitude of non-LTE effects is low.  The chemical composition
                  exhibits some underabundance of iron (a metallicity  deficit is generally observed in this object), an excess of helium,
                  and a noticeable irregularity in the
                  abundances of heavy elements. The symmetric profile of 4481 \AA\ Mg{\sc ii} line
                  is very well reproduced in the synthetic spectrum. The 9632 \AA\ line of Mg{\sc ii}  makes a major contribution
                  to a  blend with DIB9633.
  \item[\bf HD 63804] The basic stellar parameters of this object can be reliably determined based on H{\sc i} Balmer lines and on the spectral features of He{\sc i},
                  Fe{\sc i}, Fe{\sc ii}, C{\sc ii}, Si{\sc ii}. However, the
                  observed Mg{\sc ii} 4481~\AA\ line is too strong, considering any realistic abundance of magnesium.
                  Weaker Mg{\sc i} and Mg{\sc ii} lines point to  [Mg/H]=-0.3 dex. The star is probably a spectroscopic binary,
                  with components having the effective temperature of $\sim$8000 and $\sim$14000 K. The  cooler object is brighter,
                  with its spectral lines  red-shifted by $\sim$40-50 km/s. Additional spectra, acquired 1-2 years later, will permit for a more complete description.
                  Nevertheless, a  strong Mg{\sc ii}  9632 \AA\
                  line leaves  little space for the interstellar 9633~\AA\ feature. The latter has just 1/10 of the DIB9577 intensity, which is  eight times
                  weaker than expected assuming  C$_{60}^{+}$ as the carrier. The lack of weaker 9429 and 9366 \AA\ bands is obvious.
  \item[\bf HD 76341] This object is very hot O-type giant. Helium, silicon and nitrogen are in similar excess of +0.2 dex.
                  Broad stellar lines are not able to  distort the discussed infrared DIBs. This sight line offers yet another example of DIB9633 being stronger than DIB9577,
                  in contrast to the case for the laboratory gas phase spectrum of C$_{60}^+$.
  \item[\bf HD 78344] The chemical composition generally exhibits  an overabundance of helium and light elements. Broad and shallow
                   stellar lines do not affect the profiles of interstellar features.
                  The  DIB9633 intensity is just 0.6 of that found for DIB9577, instead of 0.8 expected with the assumption of C$_{60}^+$ as the carrier.
                  The absence of weaker 9429 \AA\ and 9366 \AA\ bands is evident.
  \item[\bf HD 80077] This star features a significant excess ($\sim$0.4 dex) of nitrogen, as well as of  silicon, and a deficit of carbon of almost equal magnitude.
                  Magnesium and helium abundances are almost solar. Other elements display an overabundance of $\sim$0.1 dex. The Mg{\sc ii} 9632 \AA\ line is
                  quite strong,   which  makes the true strength of
                  DIB9633  a factor of 0.6 lower than that measured for   DIB9577, instead of the value  0.8 anticipated assuming the assignments to C$_{60}^+$. The presence or absence of weaker  C$_{60}^{+}$ bands, at 9429 \AA \ and 9366 \AA,\ is doubtful owing to saturated telluric lines that significantly pollute that
                  part of the spectrum.
  \item[\bf HD136239] Excesses of +0.4 dex are found here for the alpha elements and helium.  Nitrogen and silicon are overabundant by  +0.5 dex.
                  The stellar Mg{\sc ii}
                   line is strong, making the true intensity of DIB9633  half of expected assuming   C$_{60}^+$ as the carrier. The presence/absence of 9429 \AA\ and 9366 \AA\ features is uncertain,
                  owing to saturated telluric lines.
  \item[\bf HD145502] There is only a moderate coincidence of observed and model spectra because of large differences of individual abundances
                  for different lines of helium and carbon.
                  The  deficit of carbon and magnesium is anomalous. The Mg{\sc ii}  4481~\AA\ line has an irregular shape,
                  and therefore  cannot be well reproduced by our synthetic spectrum. The star is probably a binary object.
                  A weak 9632 \AA\ Mg{\sc ii} line merges with strong He{\sc i} line to a smeared blend.
  \item[\bf HD147888] A very good match with the model spectrum is observed. The magnitude of non-LTE effects is low.
                  The chemical composition is characterized by  smoothly decreasing   elemental abundances  for heavier species.
                  Magnesium is evidently underabundant. The profile of Mg{\sc ii} 4481 \AA\ line, anomalous owing to strange, rather flat wings,
                  is nevertheless  well reproduced with the model.
                  The  Mg{\sc ii} 9632 \AA\ line makes a major contribution to the  blend with DIB9633.
  \item[\bf HD148379] The  equivalent width of DIB9633 is reduced by almost a half after  subtracting the coincident stellar magnesium line. A weak feature is possibly present
                   at 9366 \AA\ but a strong telluric contamination  prevents any precise   measurements.
  \item[\bf HD148605] There is a good coincidence of observed and model spectra. The magnitude of non-LTE effects is low.  The chemical
                  composition exhibits a smooth decrease of individual
                  abundances of chemical elements for heavier species. The profile of the Mg{\sc ii} 4481~\AA\ line is symmetric, and well
                  reproduced by the model.
                  The weak 9632\AA\ Mg{\sc ii} line merges with strong He{\sc i} line to a common smeared blend.
  \item[\bf HD167264] Again, there  is a good match of observed and model spectra. Line profiles are affected by the stellar wind.
                  The spectral features of some light elements  are perturbed by
                  uncompensated non-LTE effects. There is a  smooth decrease of individual elemental abundances for heavier species. The observed profile of Mg{\sc ii}
                  4481 \AA\ is very well reproduced by the model. There is no infrared Mg{\sc ii} line polluting DIB9633.
  \item[\bf HD168625] Here, the  intensity of DIB9633  is substantially reduced after  subtracting the contribution from Mg{\sc ii}, which makes the
                  equivalent widths ratio DIB9633/DIB9577 closer to that observed in laboratory gas phase spectra of C$_{60}^+$.
  \item[\bf HD169454] The object exhibits  almost solar abundances of helium and magnesium, while nitrogen and silicon are in moderate excess.
                   The synthetic profile of a blend made by  Mg{\sc ii} 4481 \AA\ and Al{\sc iii} 4479 \AA\   matches the observed spectrum well.
                   This sight line exemplifies the case of unusually strong 9633 \AA\ band; even after the subtraction of the stellar Mg{\sc ii}
                   feature, the equivalent width is still  higher than for DIB9577, in strong disagreement with the ratio
                   of respective bands observed in the laboratory gas phase spectrum of C$_{60}^+$.
  \item[\bf HD170740] There is a reasonable coincidence of observed and model spectra in this binary object, with evident two-component
                  profiles of He{\sc i} and C{\sc ii}. Stellar parameters were determined for the main component only. The chemical composition exhibits the deficit of
                  metallicity, with a large scatter of  abundances derived for
                  different lines of helium and carbon. Both carbon and magnesium are in deficit. The observed profile of Mg{\sc ii} 4481~\AA\
                  is almost perfectly symmetric and clearly separated
                  from the neighboring Al{\sc iii} 4479~\AA\ line, as is very well reproduced by the synthetic spectrum.
                  The intensity of DIB9633 is considerably reduced after the subtraction of  the stellar Mg{\sc ii} line contribution.
  \item[\bf HD183143] There is a good coincidence of observed and model spectra. Hydrogen lines H$_{\alpha}$ -- H$_{\delta}$ exhibit
                  strong, wind-driven emissions. The lines of other elements,
                  including neutral helium, are symmetric and match the model. The chemical composition is almost solar,
                  with a slight excess of nitrogen, probably as a result of the running CN-cycle. Mg{\sc ii} 4481\AA\ line has a symmetric profile, coincident with that generated in the synthetic spectrum.
                  A contribution from the infrared  Mg{\sc ii} 9632 \AA\ line, when subtracted, essentially reduces the intensity of DIB9633.
                  HD\,183143 is a key object for the recently announced detections of C$_{60}^+$ (Walker et al. 2015, Campbell et al. 2016).
                  As noted, this identification of interstellar C$_{60}^+$ was premature, as it laked  the analysis of the  stellar spectrum.
                  Indeed, taking account of a considerable contribution from the stellar Mg{\sc ii} line, the equivalent
                  widths ratio DIB9633/DIB9577  is just 0.4.  
  \item[\bf HD184915] There is an excellent coincidence of observed and model spectra. Non-LTE effects are moderate, without any detectable stellar
                  wind effects. The estimated rotational velocity is as high as 220 km/s.
                  The chemical composition exhibits an excessive  abundance of helium and nitrogen, probably owing to the enrichment of stellar surface
                  by CNO-cycle products. The profile of the Mg{\sc ii} 4481\AA\ line
                  is symmetric, which is well reproduced by  the model. There is no Mg{\sc ii} line overlapping DIB9633.
\end{description}

\section{Observed diffuse bands versus the C$_{60}^{+}$ laboratory gas phase spectrum}

The laboratory gas-phase absorption spectrum of C$_{60}^+$
features four main infrared bands: 9632.7, 9577.5, 9428.5 and
9365.9 \AA\ ($\pm$ 0.1 \AA), with relative intensities of 0.8,
1.0, 0.3, and 0.2, respectively; thus the strongest peak is at
9577.5~\AA\ (Campbell et al., 2015). No lines of sight in our
sample allowed the detection of that spectral pattern. In
particular, no candidates for the two weakest of the above listed
bands have emerged (cf. the second part of Table 2).

Apart from that, the assignment of diffuse bands at 9633~\AA\ and
9577~\AA\ to the two strongest C$_{60}^+$ bands  suffers from the
fact that the observed intensity ratio DIB9633/DIB9577 varies
greatly from target to target instead of scattering, within  the
known error limits, around a "canonical" value of 0.8 (see Tab.
2). Towards the majority of our targets, the 9633~\AA\ band was in
fact stronger than that at 9577~\AA. On the other hand, lines of
sight with extremely low DIB9633 intensities were revealed (Tab.
2). Such large variations of the observed strength ratio cast
serious doubts on whether the two bands could indeed share a
common carrier.

\begin{figure}
    \includegraphics[width=9 cm, angle=270]{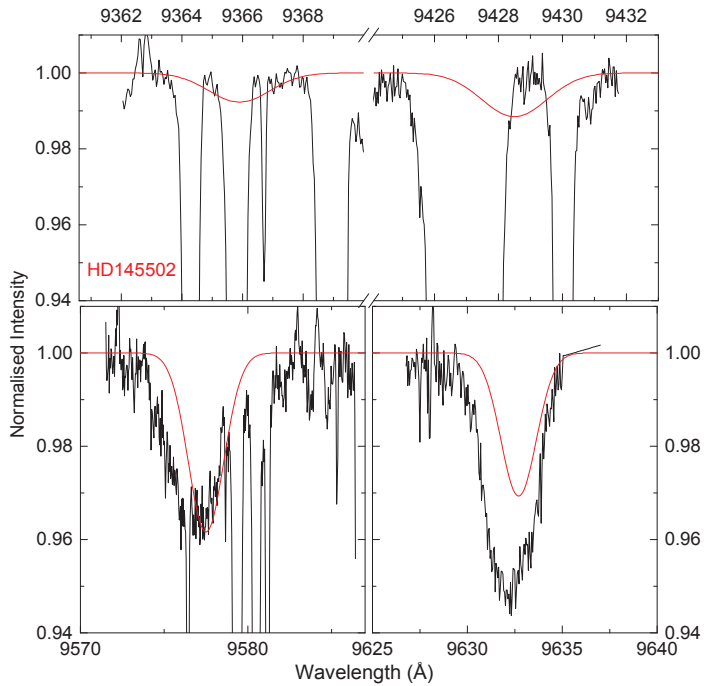} 
    \caption{Observed near infrared spectrum of HD\,145502.
    Smooth gaussian curves mimic the laboratory
    gas-phase bands of C$_{60}^+$, with wavelengths, widths, and  intensity ratios  (normalized to the
    depth of DIB9577)  taken from   the Campbell et al. (2016) study. See text for details.
     }
    \label{hd145502-4dibs}
\end{figure}

A convincing illustration of these issues is provided by the
results acquired  for HD\,145502, an object free from the Mg{\sc
ii} effect and intentionally selected to waive all speculation
concerning  the accuracy of our stellar line removal procedure.
Moreover, the elimination of telluric lines was particularly
successful for HD\,145502 (Fig. \ref{hd145502-4dibs}). For that
sight line, DIB9633 is exceptionally strong, in fact  much
stronger than DIB9577, making the DIB9633/DIB9577 ratio as big as
1.3, instead of the 0.8 observed in laboratory gas-phase spectra. Both
features are blue-shifted and  broader than expected, when
compared with the known C$_{60}^+$ bands. Concerning the two
remaining, minor bands of C$_{60}^+$, it is the case that the relevant wavelength
range is very difficult to study because of exceptionally strong
telluric lines, which are difficult to remove because of saturation effects.
Nevertheless, the upper panel of Fig. \ref{hd145502-4dibs} clearly
demonstrates that there is no room for broad and shallow features
of the expected depth. The Gaussians, normalized to the depth of DIB9577,
intersect the observed spectrum making thus the presence of both
weak features very unlikely.

The second example illustrates the opposite case, with a very
small DIB9633/DIB9577 ratio. Indeed, in the spectrum of
HD\,183143, after the Mg{\sc ii} correction, the intensity of
DIB9633 was reduced by a factor of more than 2.  The presence or
absence of weak DIB9366 cannot be confirmed owing to very strong
telluric lines which seem to be partially blended. The profile of
another weak band, DIB9428, intersects the observed spectrum in
which it can hardly be traced.

Another interesting object is HD37022. As depicted by Fig. 4,
DIB9633 is red-shifted in this spectrum, similar to what was
reported for DIB5780 (Kre{\l}owski et al. 2015). This leads to the
spectral separation of DIB9633 from the Mg{\sc ii} 9632~\AA\
line. It is the only such case in our sample. Importantly,
however, the other crucial C$_{60}^+$ candidate band, DIB9577,
which is very broad in this spectrum, does \textit{not} show any red--shift
(and may even be slightly  blue--shifted). This is further evidence
against a common origin of the two major DIBs attributed
to C$_{60}^+$.

\begin{figure}
    \includegraphics[width=9 cm, angle=270]{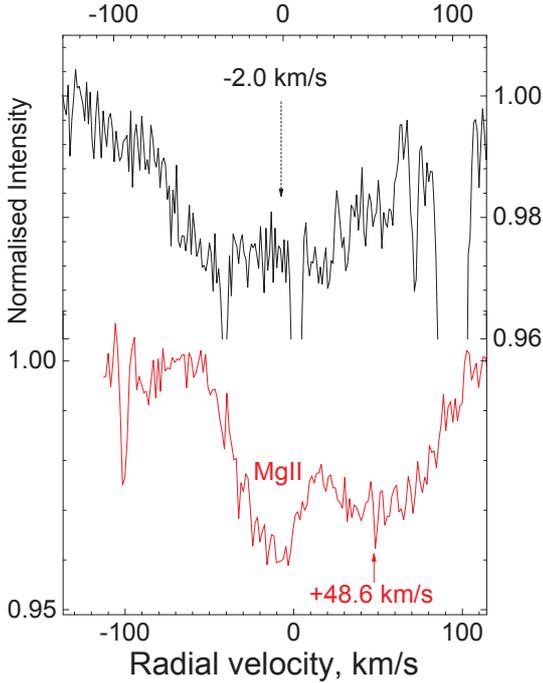}
    \caption{Radial velocity profiles for DIB9577 and DIB9633 observed towards  HD\,37022 made with "new" laboratory wavelength from Campbell et al. (2016a).
     The rest wavelength velocity scale was established using the interstellar K{\sc i} 7699\AA\ line.}
    \label{hd37022-2dibs}
\end{figure}

\begin{figure}
    \includegraphics[width=8.5 cm, angle=270]{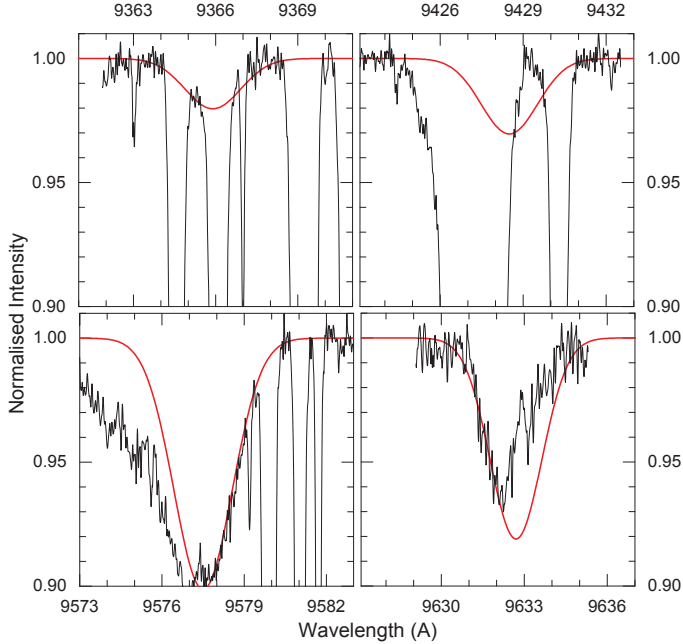} 
    \caption{Same as Fig. \ref{hd145502-4dibs}, for HD\,183143.
     }
    \label{hd183143-4dibs}
\end{figure}

Very recently Campbell et al. (2016a) have corrected laboratory
C$_{60}^+$ wavelengths, taking into account the fact that their
previously reported  values (Campbell et al. 2016) concerned weakly
bound C$_{60}^+$-He$_n$(\textit{n} = 1 - 3) complexes. The absorption
bands of such species exhibit progressive red-shifts with the
increasing number of interacting He atoms (see figs 1 and 2 in
Campbell et al., 2016a). Wavelengths estimations for bare
C$_{60}^+$ cations were based on the linear interpolation of
measurements made for the complexes containing  1, 2 and 3 He
atoms. Of note is the fact that the corrected wavelengths derived for the 2 major
peaks of bare C$_{60}^+$ (9577.0 and 9632.1 \AA, with $\pm$0.2
\AA\ as the 2$\sigma$ uncertainty) are in good agreement with the
{\bf mean} values calculated for our present  sample of  targets: 9577.0
and 9632.2 \AA. However, the observed scatter of
position of peaks in astronomical spectra greatly exceeds the uncertainty of measurements.
Diffuse bands of the common origin have to be displaced in unison, keeping the same distance
between them; in other words, the scatter of positions of diffuse bands
in astronomical measurements (Table 2) cannot be explained by an
assumption of their common origin. Indeed, as
we already reported, DIB9633 and stellar Mg{\sc ii} line at 9632
\AA\ can be easily distinguished in the spectrum of HD\,37022 (Fig.
\ref{Mg2inTwoStars}) where  DIB9633 exhibits an evident red-shift, while the
second major band 9577 fits the "new" lab wavelength well (Fig. \ref{hd37022-2dibs}). Note that in Fig.
\ref{hd37022-2dibs} both radial velocity profiles were constructed
using the "new" rest wavelengths from Campbell et al. (2016a).

Interestingly, it is not the "new", but rather "old" (i.e.
uncorrected for the He complexes) laboratory position of
the strongest C$_{60}^+$ band at 9577.5$\pm$0.1 \AA\ that
matches well the DIB9577 wavelength 9577.4$\pm$0.02 \AA\ observed towards HD\,183143 (both values are from Table 1 of Campbell et al., 2016).
Our observations of HD183143 (our Table 2) confirm the wavelength of interstellar feature at 9577 \AA.
Thus, there is surprising difference ofd 0.4-0.5 \AA\ between the "new" laboratory wavelength and that observed in HD\,183143.

Weak interactions between C$_{60}^+$ cations and some common
constituents of the interstellar gas remain to be modelled and/or
measured in laboratories. At present it's anybody's guess that such
complexes, if bound strong enough to withstand the temperatures of
the translucent interstellar medium, would have spectra differing
from that of bare C$_{60}^+$ by much more than the documented
scatter of DIB9633 and DIB9577 wavelengths/intensities. It should
also be noticed that complexes with helium, experimentally
observed by Campbell et al., formed with a high number density of
He atoms (10$^{15}$ cm$^{-3}$) and that temperatures below 8 K
were required. Such species have therefore not been postulated to
be of any significance for the DIB phenomenon.

It seems that the only astronomically observed
parameters that match the laboratory ones relatively well are
FWHMs of diffuse bands. They are slightly broader than the
laboratory ones for 9577 band, but in general the similarity is
satisfactory in most of the observed lines of sight (Table 2).

\section{Conclusions}

\begin{figure}
    \includegraphics[width=4.5 cm, angle=270]{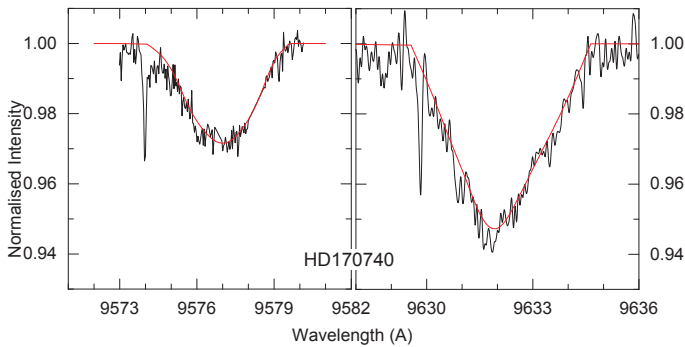}
    \caption{Diffuse bands at 9577 and 9633~\AA\ (corrected profiles, same ordinate scale). The latter is 1.6 times stronger than the former, while the respective ratio, expected for C$_{60}^+$, is 0.8! Note that that the originaly observed intensity ratio, before the subtraction of a stellar Mg{\sc ii} contribution, was even greater, as big as 1.9.}
    \label{hd170740}
\end{figure}

\begin{figure}
    \includegraphics[width=7 cm, angle=270]{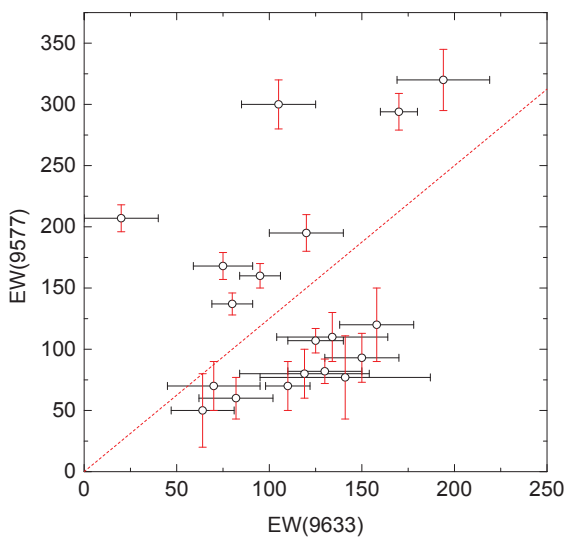}
    \caption{Very poor intensity correlation of 9633 and 9577~\AA\ diffuse bands. Dotted line represents the equivalent width ratio DIB9633/DIB9577=0.8. Differences in strength ratios are much bigger than measurement errors.}
    \label{dibs9633vs9577}
\end{figure}

The results from the observed sample of 19 reddened stars having spectra with
easily detectable diffuse bands at 9577 and 9633~\AA\ do not
allow us to assign these features to near-IR transitions of
C$_{60}^{+}$, and thus to confirm the presence of this cation in
translucent interstellar clouds, for the following reasons:

\begin{itemize}
  \item{The ratio of DIB9633 and DIB9577 equivalent widths is variable within  a broad range: see Table 2 and, for example, Figs. \ref{hd145502-4dibs}, \ref{hd183143-4dibs}, \ref{hd170740}.
        The resultant mutual correlation is very poor, much worse than within any other pair of reasonably strong DIBs (Fig. \ref{dibs9633vs9577}).
        }
  \item{
       We have not confirmed the presence of two weaker members of the C$_{60}^+$ family, expected at 9428 and 9366~\AA\; the former is stronger in laboratory but more difficult to be traced in observations, even though these should appear assuming the validity of experimentally determined intensity ratios.
       }
  \item{The ``interstellar'' wavelengths of the two strongest DIBs proposed for C$_{60}^+$ show evident variability, but not in unison, i.e. the distance between DIB9633 and DIB9577 discernibly fluctuates.
       }
  \item{The profile shapes of both diffuse bands are variable. In particular, asymmetric profiles were found in several cases, contrasting with what was reported for gas-phase C$_{60}^+$.}
\end{itemize}

In general, the observed very poor correlation of the two discussed DIBs can be explained, together with the variable shapes of DIB profiles, by some blending with other,
as yet unresolved diffuse bands. Nevertheless, both the lack of a firm detection of the two minor bands expected for C$_{60}^+$ and, for the major candidate bands,
the disagreement of central wavelengths with laboratory values need to be addressed in order to defend the recently claimed identification of C$_{60}^+$ .

Cami et al. (2010) suggested that ``the absence of the
corresponding spectral features of fullerene cations and anions
implies that the fullerenes are in the neutral state''. This
remark, formulated in the context of {\it circumstellar} matter,
may prove correct for translucent interstellar clouds also.

\section*{Acknowledgements}

This paper includes data gathered with the VLT and UVES spectrograph, programs 067.C-0281(A), 082.C-0566(A), 092.C-0019(A).
Authors acknowledge Dr. H. Linnartz and Dr. R.Kolos for his valuable comments and suggestions. GAG and GV acknowledge the support of Russian
Science Foundation (project 14-50-00043, area of focus Exoplanets). VVS acknowledges the Russian Fund for Basic Researches
16-02-01145 (Non-LTE modeling of stellar atmospheres). JK acknowledges the grant 2015/17/B/ST9/03397 of the Polish National Science Center.

\bsp    
\label{lastpage}
\end{document}